\documentstyle[aps,amssymb]{revtex}

\begin{document}
\title{Field-induced vacancy localization in a driven lattice gas: Scaling of
steady states. }
\author{M. Thies$^{1,2}$ and B. Schmittmann$^{2}$}
\address{$^{1}$Lehrstuhl WTM, Universit\"{a}t Erlangen-N\"{u}rnberg, 
Martensstr. 5, 91058 Erlangen, Germany;\\
$^{2}$Center for Stochastic Processes in Science and Engineering and
Department of Physics, \\
Virginia Tech, Blacksburg, Va 24061-0435, USA.}
\date{\today}

\maketitle

\begin{abstract}
With the help of Monte Carlo simulations and a mean-field theory, we
investigate the ordered steady-state structures resulting from the motion of
a single vacancy on a periodic lattice which is filled with two species of
oppositely ``charged'' particles. An external field biases particle-vacancy
exchanges according to the particle's charge, subject to an excluded volume
constraint. The steady state exhibits charge segregation, and the vacancy is
localized at one of the two characteristic interfaces. Charge and hole
density profiles, an appropriate order parameter and the interfacial regions
themselves exhibit characteristic scaling properties with system size and
field strength. The lattice spacing is found to play a significant role
within the mean-field theory.
\end{abstract}

\section{Introduction}

Systems in non--equilibrium steady states have attracted considerable
interest in the past decade \cite{schmittmannbook}. While presenting a
wealth of unexpected, intriguing phenomena, they are still quite poorly
understood at a fundamental level. It is therefore natural to investigate
simple model systems to identify generic behavior, before turning to real
systems which are usually far more complex.

A particularly interesting class of model systems is based on lattice gas
models, involving one or several species of particles whose motion is biased
in a specified direction. If the boundary conditions are open or periodic,
the bias can drive the system out of a well-known equilibrium state, into
novel non-equilibrium steady states which typically carry global particle
currents. Characteristic configurations, particle-particle correlations and
even phase transitions tend to be profoundly affected by the bias.
Equilibrium phases can be suppressed, universality classes may change and
entirely new transitions can emerge. For example, in a simple driven Ising
lattice gas with periodic boundary conditions \cite{katz}, the bias
suppresses one of the two ground states of the equilibrium system and
fundamentally changes the universal properties of the Ising order-disorder
transition \cite{newuniversalityclasses,leung2,schmittmann1991,pz}. In the
high-temperature phase, it induces generic long-range correlations \cite
{farfromtc}, which characterize all models of this type. Other anomalies are
observed below $T_{c}$ \cite{lowT}.

If the Ising symmetry is generalized \cite{potts,blume} to include two (or
more) species of particles which respond differently to the drive, such
systems will generically exhibit blocking transitions, similar to traffic
jams, in which one species impedes the motion of the other. These
instabilities are genuine non-equilibrium transitions: they do not exist in
the equilibrium limit and are controlled by particle density and bias
strength, rather than temperature. The ordered phases exhibit characteristic
spatial structures. Related real and model systems include water-in-oil
microemulsions in external electric fields \cite{chandra,waterdroplets}, gel
electrophoresis \cite{gelelectro} and traffic flow \cite{trafficflow}.

In this paper, we focus on a three-state lattice gas consisting of holes and
two distinct species of particles driven by an external field $E$ (the bias) in
opposite directions \cite{schmittmann1992}. We name the two species
``positive'' and ``negative'', in analogy to charged particles in an
electric field. The bias clearly breaks the Potts symmetry of the stochastic
variable, by acting differently on each species. The only interaction
between the particles is an excluded volume constraint so that (i) each site
can be occupied by at most one particle, and (ii) particle-particle
(``charge'') exchanges are not allowed. In the absence of other
interparticle interactions, the temperature dependence of the system,
reflecting a coupling to a heat bath, may be absorbed into the drive. Hence,
the model is a high--field, high--temperature limit of a more complicated
interacting system.

On a fully periodic square lattice, this system undergoes a blocking
transition controlled by field strength and particle density, separating a
homogeneous phase from a spatially inhomogeneous one \cite{schmittmann1992}.
For small mass density and drive, the steady--state configurations are
disordered, so that both particle densities are homogeneous and a
significant charge current persists. In contrast, if a threshold mass
density is exceeded, the particles form a single compact strip transverse to
the field while the rest of the lattice remains essentially empty. The
particle-rich region itself consists of two strips, also oriented transverse
to the field, each dominated by one single species. In this phase, the
particles impede one another, due to the excluded volume constraint, so that
the charge current is much smaller. Other ordered phases, with nonzero
winding number around the lattice (``barber poles''), are observed in
systems with rectangular aspect ratios \cite{bassler}. An analytical
solution in the frame of a mean-field theory \cite{schmittmann1992} was
presented in Ref.~\cite{vilfan}. With a slight geometrical modification, the
model was also investigated by Foster and Godr\`{e}che \cite{godeche}.

Here, we focus on a novel aspect of the blocking transition, namely, a{\em \
localization phenomenon} occurring in systems near complete filling \cite
{korniss}. Thus, all lattice sites except a {\em single} vacant one are
occupied by particles. For simplicity, we consider the symmetric situation,
i.e., the particle numbers of each species differ at most by one. Starting
from a disordered initial configuration, particles may exchange only with
the vacancy. As a result, the hole diffuses through the lattice. However, it
does {\em not} perform a Brownian random walk, since the jump rate for a
particle-hole exchange depends on the charge and direction of motion of the
particle. By virtue of the bias, positive and negative particles are
transported in opposite directions: The two particle species eventually
segregate, provided the field exceeds a certain threshold, corresponding to
the transition line \cite{schmittmann1992}. When the steady state is
reached, two strips have formed, filled by positive and negative particles,
respectively. The hole itself ends up ``trapped'' on one of the two
interfaces between the two ordered regions. Its location is the remnant of
the empty region observed at finite hole density.

This problem, in both its static and dynamic aspects, is an example of a
much-wider ranging class of interacting random walk and defect-mediated
domain growth problems. The hole is a random walker whose motion changes its
environment, but the environment reacts by determining the local jump rates.
The vacancy plays the role of a highly mobile defect \cite{zoltan},
interacting with an otherwise immobile background. The time evolution of the
system, from an initially disordered particle background to two ordered
strips, poses a domain growth problem \cite{bray}. Clearly, a good
understanding of the final {\em steady states} and their associated scaling
properties is the first step in the analysis of the ordering process. This
study forms the subject of this paper. We report elsewhere on the full {\em %
dynamics} \cite{dynamics}.

The key results of our study \cite{master} can be summarized as follows.
First, we establish and confirm the characteristic scaling forms of the
order parameter and the density profiles. Further, focusing only on the
interfacial (as opposed to the fully ordered) regions of the profiles, we
find that {\em both} interfaces are independent of the longitudinal system 
size and that their
widths are controlled by the drive alone. 
These findings are reflected in our mean-field
theory. Our results are limited in two ways: first, by the onset of the
phase transition for small $E$, and second, by the breakdown of the naive
continuum limit for large $E$.

This paper is organized as follows. In Section 2, we give a precise
definition of the microscopic model which underlies the Monte Carlo
simulations. The relevant control and order parameters are defined. To set
the scene, we provide a brief summary of earlier work. In particular, we
discuss the blocking transition and its description in terms of a mean-field
theory. In Section 3, we investigate the scaling properties of the order
parameter and the profiles, based on Monte Carlo simulations and the exact
solution of the mean-field equations. We conclude with a summary and some
comments.

\section{The Model: Microscopics and Mean-Field Theory}

\label{mcsimul}In this section, we provide the necessary background. We
begin with the microscopic definition of the model, followed by a summary of
its phenomenology. We then provide a different perspective, by sketching the
mean-field theory and its main results. We close with some technical details
of the simulations.

Our model is defined on a two--dimensional square lattice of $L_{x}\times
L_{y}$ sites with fully periodic boundary conditions. Each site, except one,
can be occupied by a positive or negative particle. The remaining site is
left empty. The resulting configurations can be described by an occupation
variable $n_{xy}^{+}$ ($n_{xy}^{-}$), taking the value $+1$ if a
positive (negative) charge is present at site ($x,y$) and zero otherwise.
This enforces the excluded volume constraint. There are no other
interactions between the particles. Turning to dynamics, particles may jump 
{\em only} onto the vacant site. In the absence of the external field, the
vacancy exchanges randomly with any of its four nearest neighbors,
independent of their charge and the direction of the move. This symmetry is
broken by the ``electric'' field $E$, which is chosen to be uniform in space
and time and directed along the positive $y$--axis. For nonzero $E$, jumps
transverse to the field are still random; however, parallel jumps are now
biased: positive (negative) charges jump preferentially along (against) $E$.
Specifically, the exchange rate of the hole with a randomly chosen nearest
neighbor is given by the Metropolis rate \cite{metropolis}: 
\begin{equation}
W=\min \left\{ 1,\exp (qE\delta y)\right\}   \label{transitionrate}
\end{equation}
where $q=+1$ ($-1$) for a positive (negative) particle and $\delta y=0,\pm a$ is the
change of the $y$--coordinate of the particle due to the jump. This choice
mimics the local energetics of charges in a uniform field. The lattice
constant $a$ will be set to $1$.

The dynamics of the model can be summarized by a master equation \cite
{bindermcs}, for the probability $P(C,t)$ to find the system in the
configuration $C=\left\{ n_{xy}^{+},n_{xy}^{-}\right\} $ at time $t$: 
\begin{equation}
\frac{\partial }{\partial t}P(C,t)=\sum_{C^{\prime }}\left\{ W(C^{\prime
}\rightarrow C)P(C^{\prime },t)-W(C\rightarrow C^{\prime })P(C,t)\right\} .
\label{masterequ}
\end{equation}
Here, $W(C\rightarrow C^{\prime })$ is the transition rate from $C$ to $%
C^{\prime }$, specified by Eqn (\ref{transitionrate}). For $E<\infty $, $%
P(C,t)$ approaches a unique steady-state solution $P^{*}(C)$ in the limit $%
t\rightarrow \infty $. For closed boundary conditions, $P^{*}(C)$ follows
from equilibrium statistical mechanics, being the Boltzmann factor of a
system of noninteracting charges in a uniform field. For periodic boundary
conditions, however, there is no uniquely defined static potential for $%
E\neq 0$ so that $P^{*}(C)$ is not a priori known. Instead, it has to be
found from an explicit solution of Eqn (\ref{transitionrate}).
Unfortunately, such solutions are available only for a few, mostly
one-dimensional, cases. Here, we only know the $E=0$ solution: the system is
again in equilibrium, the particles diffuse randomly and $P^{*}(C)$ is
independent of configuration, i.e., $P^{*}\propto 1$.

The control parameters of this model are easily identified. In addition to
the driving field $E$ and the system size, $L_{x}\times L_{y}$, we can
adjust the mass density 
\begin{equation}
m\equiv \frac{1}{L_{x}L_{y}}\sum_{x,y}\left( n_{xy}^{+}+n_{xy}^{-}\right)
\label{mass}
\end{equation}
as well as the net charge density of the system: 
\begin{equation}
\rho \equiv \frac{1}{L_{x}L_{y}}\sum_{x,y}\left(
n_{xy}^{+}-n_{xy}^{-}\right) \quad \text{.}  \label{charge}
\end{equation}
Since the particle number of each species is separately conserved, both
densities are also conserved. For our case, there is always a single hole, so
that the particle density 
\begin{equation}
m=1-\frac{1}{L_{x}L_{y}}  \label{1hole}
\end{equation}
depends on the system size. Since we focus on nearly equal numbers of
positive and negative particles, the net charge density is zero for systems
with an odd number of sites and $-1/(L_{x}L_{y})$ for an even number (the
hole always takes the place of a positive particle). In the simulations,
this small difference does not appear to lead to observable effects, unlike
the case of $\rho =O(1)$ \cite{LZ}.

A brief description of the blocking transition and the associated phases
will be helpful. For small values of drive and total mass, the system is in
the disordered phase, characterized by spatially uniform mass and charge
densities. A significant charge current flows in this phase. As $E$ or $m$
increase, a transition into an ordered phase, with spatially inhomogeneous
densities, occurs. For systems with aspect ratios near unity, each species
of particles forms a compact, stable strip transverse to the electric field.
The strip of positive charges is located directly ``upfield'' from the
negative strip, so that the strips block each other, due to the excluded
volume constraint. The rest of the lattice remains essentially empty.
Clearly, the current is much smaller in this phase. In the following, we
will investigate the structure of these transverse strips when the empty
region has shrunk to a single hole. We never observe strips with nonzero
winding number: they appear to be suppressed near complete filling.

To distinguish ordered and disordered phases, a suitable order parameter is
needed. It is convenient to introduce the local {\em hole} and {\em charge}
densities: 
\begin{equation}
\phi _{x,y}=1-\left( n_{x,y}^{+}+n_{x,y}^{-}\right) \qquad \text{and}\qquad
\psi _{x,y}=n_{x,y}^{+}-n_{x,y}^{-}.  \label{local}
\end{equation}
Since our system does not develop inhomogeneities in the $x$--direction, it
is natural to focus on the mass and charge density {\em profiles}: 
\begin{equation}
\phi (y)=\frac{1}{L_{x}}\sum_{x}\phi _{x,y}\qquad \text{and}\qquad \psi (y)=%
\frac{1}{L_{x}}\sum_{x}\psi _{x,y}.  \label{profiles}
\end{equation}
Following Ref.~\cite{schmittmann1992}, we define an order parameter 
\begin{equation}
Q_{L}\equiv \frac{1}{m\,L_{y}}\left\langle \sum_{y}\left( \psi (y)\right)
^{2}\right\rangle \quad .  \label{ql}
\end{equation}
The angular bracket denotes a configurational average. Squaring the charge
density profile (which can have either sign) prevents unwanted cancellations
in the sum over $y$. In the ordered phase, $Q_{L}$ is $O(1)$, while being
only order $O(1/(mL_{x}))$ in the disordered phase. Roughly speaking, $%
m\,L_{y}Q_{L}$ counts the ordered rows transverse to the external field. For
a perfectly ordered system, $Q_{L}$ would be unity. Clearly, other
definitions of an order parameter are possible. In particular, the amplitude
of the lowest Fourier component of either $\psi (y)$ or $\phi (y)$ is a much
more sensitive measure for a study of the transition line \cite
{leung2,korniss}. Here, however, our focus is not on the transition, but on
the structure of ordered states, so that $Q_{L}$ serves its purpose well.

Finally, let us add a comment on the transition line. Earlier simulation
data \cite{schmittmann1992} show that the threshold mass, $m_{c}$, depends
strongly on $E$ and the longitudinal system length $L_{y}$, but only weakly
(if at all) on $L_{x}$. It can be first or second order \cite{vilfan,korniss}%
, in different regions of parameter space. In our case, where only a single
hole is present, the mass density is unity, to excellent accuracy.
Therefore, the transition is controlled by $E$ and $L_{y}$ alone.

We now turn to a brief summary of the theoretical analysis \cite
{schmittmann1992,vilfan} which will be essential for the following. Even
though the master equation is just a linear equation, in practice it is not
susceptible to theoretical analysis. To proceed, a continuum description is
introduced, in the form of equations of motion for the {\em coarse-grained}
hole and charge density profiles. Since the latter are both conserved
quantities, the equations of motion take the form of continuity equations.
They can be derived phenomenologically \cite{schmittmann1992} or directly
from the master equation \cite{korniss}. In the latter case, we first write
a set of equations for the local averages, $\left\langle \phi
_{x,y}\right\rangle $ and $\left\langle \psi _{x,y}\right\rangle $ on
discrete space (with lattice constant $1$) and then take a {\em naive}
continuum limit, e.g., we approximate $\left\langle \frac{1}{2}(\phi
_{x+1,y}-\phi _{x-1,y})\right\rangle $ by a first derivative with respect to 
$x$, etc. A mean-field assumption is necessary since two-point correlations
must be truncated in order to obtain a closed set of equations. These can
easily be written in general dimension $d$: 
\begin{eqnarray}
\partial _{t}\phi (\vec{r},t) &=&\vec{\nabla}\cdot \left\{ \vec{\nabla}\phi +%
{\cal E}\,\phi \psi \,\hat{y}\right\}   \nonumber \\
\partial _{t}\psi (\vec{r},t) &=&\vec{\nabla}\cdot \left\{ \phi \,\vec{\nabla%
}\psi -\psi \,\vec{\nabla}\phi -{\cal E}\,\phi (1-\phi )\,\hat{y}\right\} ,
\label{MFT}
\end{eqnarray}
Here, the hole density $\phi $ and the charge density $\psi $ are functions
of the $d$-dimensional coordinate $\vec{r}$ (with associated gradient $\vec{%
\nabla}$), and time $t$. The drive appears in these equations via its
coarse-grained equivalent, the {\em effective} drive ${\cal E}$: 
\begin{equation}
{\cal E}(E)=2\,\tanh (E/2).  \label{curly}
\end{equation}
pointing along unit vector $\hat{y}$. A diffusion constant has been absorbed
into the time scale. Derivatives higher than second order have been
neglected, anticipating smoothly varying solutions. The equations have to be
supplemented with periodic boundary conditions and the constraints on total
mass and charge. For later reference, we also define the parameter 
\begin{equation}
\epsilon \equiv {\cal E}L_{y}  \label{eps}
\end{equation}
which will play the role of a scaling variable.

Time-independent solutions of these equations reflect stationary phases of
the discrete model. The disordered phase corresponds to a homogeneous
solution, which is stable with respect to small perturbations provided $m$
does not exceed a threshold value $m_{H}$, given by 
\begin{equation}
m_{H}=[1+(2\pi /\epsilon )^{2}]/2\quad .\quad  \label{mh}
\end{equation}
The profiles in this phase are uniform. In our case, where the lattice is
nearly completely filled, i.e., $m\lesssim 1$, we need $\epsilon \lesssim
2\pi $ in order to find a stable homogeneous steady state. For an electric
field $E=1.0$, this implies rather small system sizes ($L_{y}<7$).

To find a steady-state solution which corresponds to a transverse strip, we
seek solutions that are inhomogeneous in the $y$-coordinate only. Eqns~(\ref
{MFT}) can be integrated once, with integration constants being the hole and
the charge currents. The former vanishes by symmetry at zero total charge.
The latter, being non--zero in general, will be denoted by $j{\cal E}$.
After expressing $\psi $ in terms of $\phi $: 
\begin{equation}
\psi (y)=\frac{\phi ^{\prime }(y)}{{\cal E}\phi (y)}  \label{psid}
\end{equation}
and rescaling the spatial variable to $z\equiv y/L_{y}$, we obtain an
ordinary differential equation, for the function $\chi \equiv 1/\phi $: 
\begin{equation}
\chi ^{\prime \prime }(z)/\epsilon ^{2}=-j\chi ^{2}(z)+\chi (z)-1.
\label{de1}
\end{equation}
To satisfy the boundary conditions, $\chi $ should be periodic with period $%
1 $. Writing $\chi $ in terms of a potential $(1/\epsilon ^{2})\chi ^{\prime
\prime }=-(d/d\chi )\,V(\chi )$, a further integration leads to $\chi
^{\prime }=\epsilon \sqrt{2(U-V)}$, where $U$ is another integration
constant. Unique solutions exist for $j<1/4$ and appropriate $U$.
Introducing the three roots $\chi _{1}\le \chi _{-}\le \chi _{+}$, defined
via $2(U-V(\chi ))=(2j/3)\,(\chi _{+}-\chi )(\chi -\chi _{-})(\chi -\chi
_{1})$, the solution \cite{vilfan} can be written using Jacobian elliptic
functions \cite{abramowitz}: 
\begin{equation}
\chi (z)=\chi _{+}-(\chi _{+}-\chi _{-})\,\text{sn}^{2}\left( \epsilon z%
\sqrt{(j/6)\,(\chi _{+}-\chi _{1})}\right) .  \label{chi}
\end{equation}
in the interval $0\le z\le 1/2$. The other half of the interval, $1/2\le
z\le 1$, is described by symmetry around the point $z=1/2$. Thus, the hole
density takes its {\em minimum} at $\phi (0)=1/\chi _{+}$, and its maximum
at $\phi (\frac{1}{2})=1/\chi _{-}$. The third root, $\chi _{1}$, lies
outside the physical domain. It is convenient to define the parameters $p$
and $R$: 
\begin{eqnarray}
p &\equiv &(\chi _{+}-\chi _{-})/(\chi _{+}-\chi _{1})  \label{p} \\
R &\equiv &\left( 4K(p)/\epsilon \right) ^{2}.  \label{r}
\end{eqnarray}
Here, $K$ stands for the complete elliptic integral \cite{abramowitz} and is
a function of $p$. Quantities of interest, such as the mass $m$ or current $%
j $, can be expressed in terms of $p$ and $R$: 
\begin{eqnarray}
1-4j &=&R^{2}\,(1-p+p^{2})  \label{1-4j} \\
1-m &=&\frac{\left[ 1-R^{2}(1-p+p^{2})\right] \,\Pi (n|p)}{2(1+R+pR)\,K(p)},
\label{1-m}
\end{eqnarray}
where $\Pi (n|p)$ is the complete elliptic integral of the third kind and $%
n\equiv 3pR/(1+R+pR)$. In principle, Eqns (\ref{p}) and (\ref{1-m}) can be
inverted to give the physical parameters $m$ and $\epsilon $ in terms of $R$
and $p$. In practice, it is easier to generate functions of interest, e.g. $%
j(\epsilon ,m)$ or the order parameter $Q_{L}(\epsilon ,m)$, parametrically
in $p$, which is allowed to range from $0$ to an upper limit $p_{0}(\epsilon
)<1$ \cite{vilfan}. The upper limit $p_{0}$ is defined by the vanishing of
the current, $j(\epsilon ,p_{0})=0$, and plays a particularly important role
in the context of this study: According to Eqn~(\ref{1-m}), the mass density 
$m$ tends to unity as $p$ approaches its upper limit $p_{0}(\epsilon )$.
Thus, only values of $p$ near $p_{0}$ will be of interest here, since our
focus is on nearly filled systems. This observation is used later for
approximations.

The solution for $\chi (z)$ generates both, hole and charge, densities: $%
\phi =1/\chi (z)$ and $\psi =\chi ^{\prime }/(\epsilon \chi )$. These
solutions describe the ordered phase, i.e., particle-rich strips transverse
to the field. For {\em fixed} mass, they depend only on the parameter $%
\epsilon ={\cal E}\,L_{y}$ and the variable $z=y/L_{y}$; thus, these
functions satisfy scaling in these variables. Moreover, the order parameter $%
Q_{L}$ is a function of $\epsilon $ alone, since the spatial variable is
integrated out. Here, however, we have to be rather more careful: since our
system, irrespective of its size, will always contain only a single hole,
the mass is inherently size-dependent. We will return to this issue in the
next section.

Since it is cumbersome to work with Eqn~(\ref{chi}) directly, its
approximation for $\epsilon \gg 1$ is very useful \cite{vilfan}. The
sn--function can be replaced by a $\tanh $--function, and the argument
simplifies : 
\begin{equation}
\chi (z)\simeq \chi _{+}-(\chi _{+}-\chi _{-})\,\tanh ^{2}\left( \epsilon
z/2\right) .  \label{chiappr}
\end{equation}
As a result, a (weak) discontinuity appears in the first derivative of $\chi 
$ at the symmetry point $z=1/2$. This is unfortunate for our purposes, since 
$z=1/2$ is also the location of the maximum hole density. A different
approximation, to be presented in the next section, resolves this
difficulty. We note in passing that Eqn (\ref{chiappr}) takes the form of
the soliton in the Korteweg--de Vries equation \cite{KdV}.

Clearly, one should not expect such a mean-field theory to provide a
quantitatively correct description of the phase transition. However, it
gives excellent {\em qualitative} insight into the instability and the phase
diagram \cite{vilfan}. Moreover, since our interest here focuses on behavior 
{\em deeply} in the ordered phase, fluctuations do not play a significant
role, and a mean-field theory should be very reliable. In fact, we will see
that its main limitations do not arise from the neglect of correlations, but
from taking a naive continuum limit.

We conclude this section with a few technical details of the simulations.
The linear system sizes, $L_{x}$ and $L_{y}$, range from $16$ to $48$, with $%
E$ in the range $0.2$ to $1.2$. A characteristic parameter set is that of
our ``reference system'', which will appear in all scaling plots: $E=0.8$
and $L_{x}\times L_{y}=16\times 24$. Thus, the mass density differs from
unity by at most $0.4\%$, so that $m=1$ is often an excellent approximation.
The statistical error of the simulation results is of the order of $5\%$ and
thus much larger. All initial configurations are random. In one MCS, a
nearest neighbor of the vacancy is chosen at random and an exchange is
attempted with the rates (\ref{transitionrate}). Averages are computed from $%
100$ independent samples for each choice of parameters. The approach to
steady state is extremely slow \cite{dynamics} for larger system sizes and
sets real-time limits on our simulations. For example, a system with $%
16\times 24$ sites at $E=0.8$ requires approximately $5\times 10^{7}$ MCS to
reach the steady state. If we increase $L_{y}$ from $24$ to $36$, which is a
factor of $1.5$ in system size, the required number of MCS increases by
roughly a factor of $10$.

While averaging, e.g., $Q_{L}$, is rather simple, by first measuring $Q_{L}$
for each sample and then averaging these data, some effort is needed to
compute {\em average} density {\em profiles} from the configurational data.
Due to translational invariance, strips can be centered at any $y$, and a
careless average would ``wash out'' any inhomogeneities. To avoid this, we
first shift the ordered strips in the different samples in such a way that
they match before we average. A natural choice would be to center all strips
on, e.g., $y=0$, by normalizing the phase of the largest wavelength Fourier
component of the profile \cite{korniss}. This is particularly useful when
profiles are measured near the phase transition. Here, however, we will
mostly take data deeply in the ordered phase, where the hole is essentially
trapped. Thus, for each sample, we keep track of the $y$--position of
the hole and determine the maximum of the hole density after a large number
of MCS. This maximum marks the interface between the positively and
negatively dominated regions, the former located ``up-field'' from the
latter. The charge density profiles from different samples are now shifted
such that these maxima coincide, and averages can be taken. Clearly,
this procedure would run into difficulties if the interface were to wander
significantly while the data for the hole density profile are being
accumulated. However, for the choices of the control parameters considered
here, this does not appear to present major problems since fluctuations of
the interface position are rather small. Moreover, they are very slow; thus,
the time scales over which the interface remains well localized are
sufficiently large to determine the maximum of the hole density very
precisely.

\section{Scaling Behavior in the Steady State}

As an introduction to the discussion of scaling properties, we illustrate
the process by which the system approaches the steady state. A series of
snap shots, taken at different MC times, demonstrates, first, why the
dynamics is so slow, and second, already suggests one of the key hypotheses
of this work, namely, that the steady-state interfaces are well separated from one
another. Figure 1 shows this series for our reference system. The negative
(positive) particles are colored black (white) and the empty site is marked
gray. A coordinate system is introduced in the usual way, i.e., the $x$%
--direction lies horizontal, the $y$--direction vertical and the $E$--field
points upwards.

Starting from a random configuration (Fig.~1a), the system remains
disordered for early times (Fig.~1b). Eventually, by allowing positive
(negative) particles to move preferentially upwards (downwards), the hole
begins to segregate the two species. The early stages of this process are
discernible in 
Fig.~1c, where an interface between regions of opposite charge begins to
develop. The position of this interface is determined by random fluctuations 
in the system. Clearly, 
\begin{figure}[tbp]
 \input epsf           
  \hfill\hfill
\begin{minipage}{0.16\textwidth}
  \epsfxsize = \textwidth \epsfysize = 1.5\textwidth \hfill
  \epsfbox{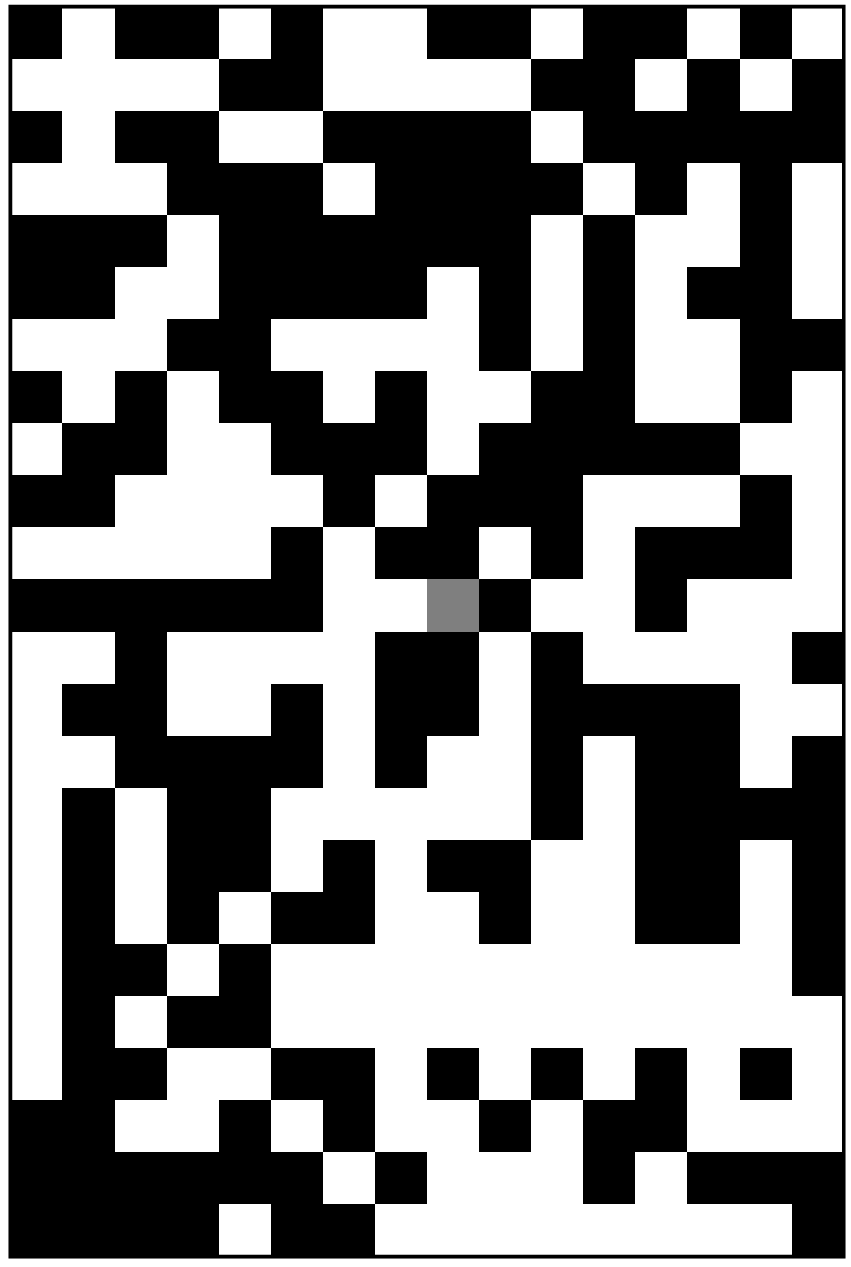} \hfill
    \begin{center} a) \end{center}
  \end{minipage}
  \hfill \hfill
\begin{minipage}{0.16\textwidth}
  \epsfxsize = \textwidth \epsfysize = 1.5\textwidth \hfill
  \epsfbox{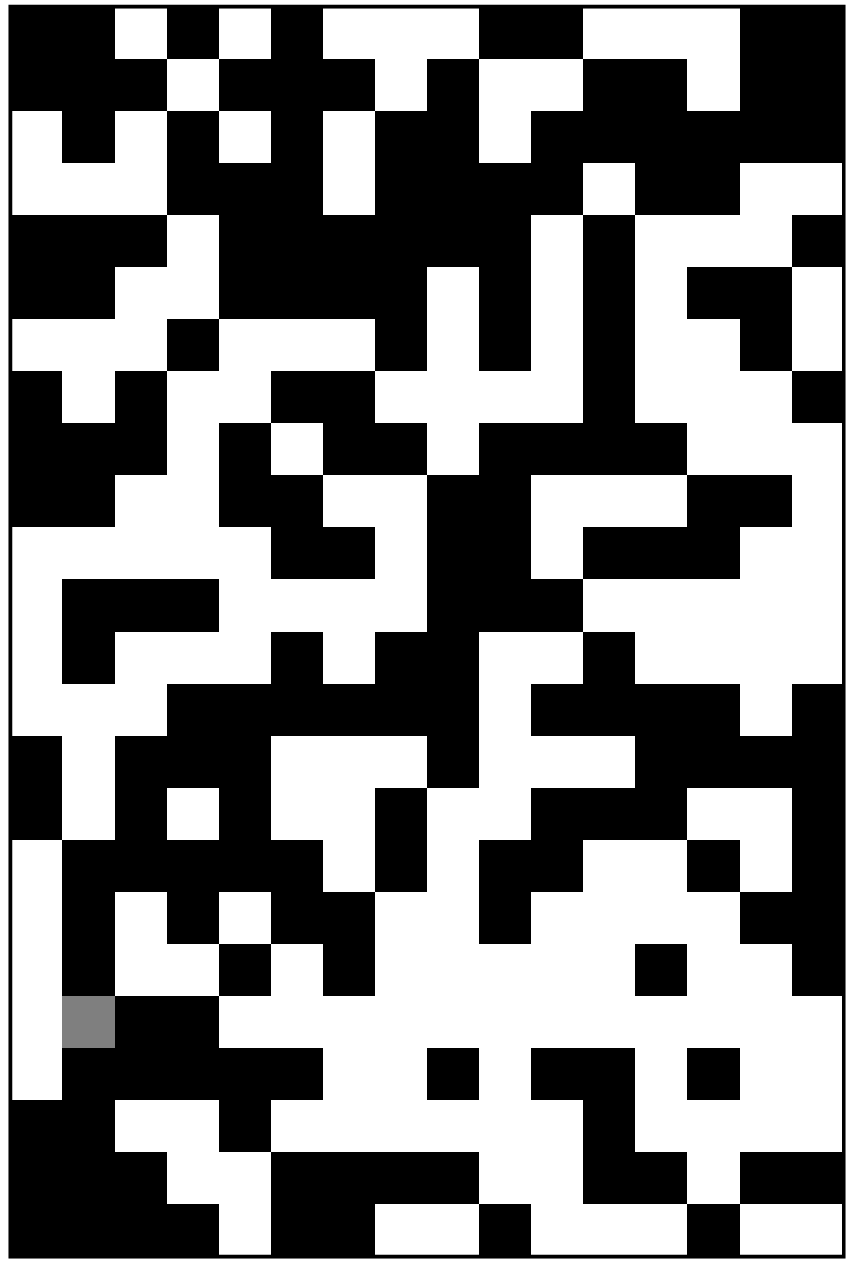} \hfill
    \begin{center} b) \end{center}
  \end{minipage}
  \hfill \hfill
\begin{minipage}{0.16\textwidth}
  \epsfxsize = \textwidth \epsfysize = 1.5\textwidth \hfill
  \epsfbox{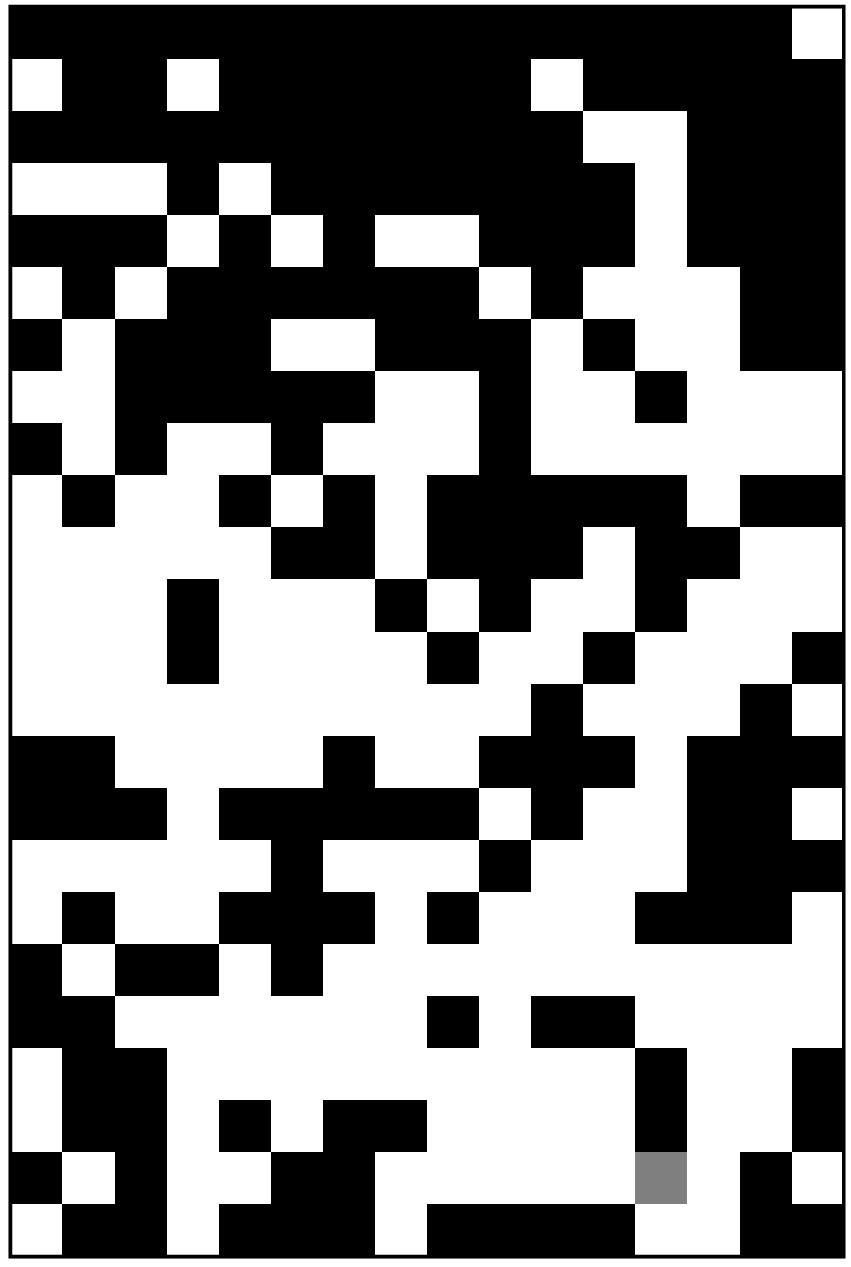} \hfill
    \begin{center} c) \end{center}
  \end{minipage}
  \hfill\hfill \vspace{0.02\textwidth}
\par
\hfill\hfill
\begin{minipage}{0.16\textwidth}
  \epsfxsize = \textwidth \epsfysize = 1.5\textwidth \hfill 
  \epsfbox{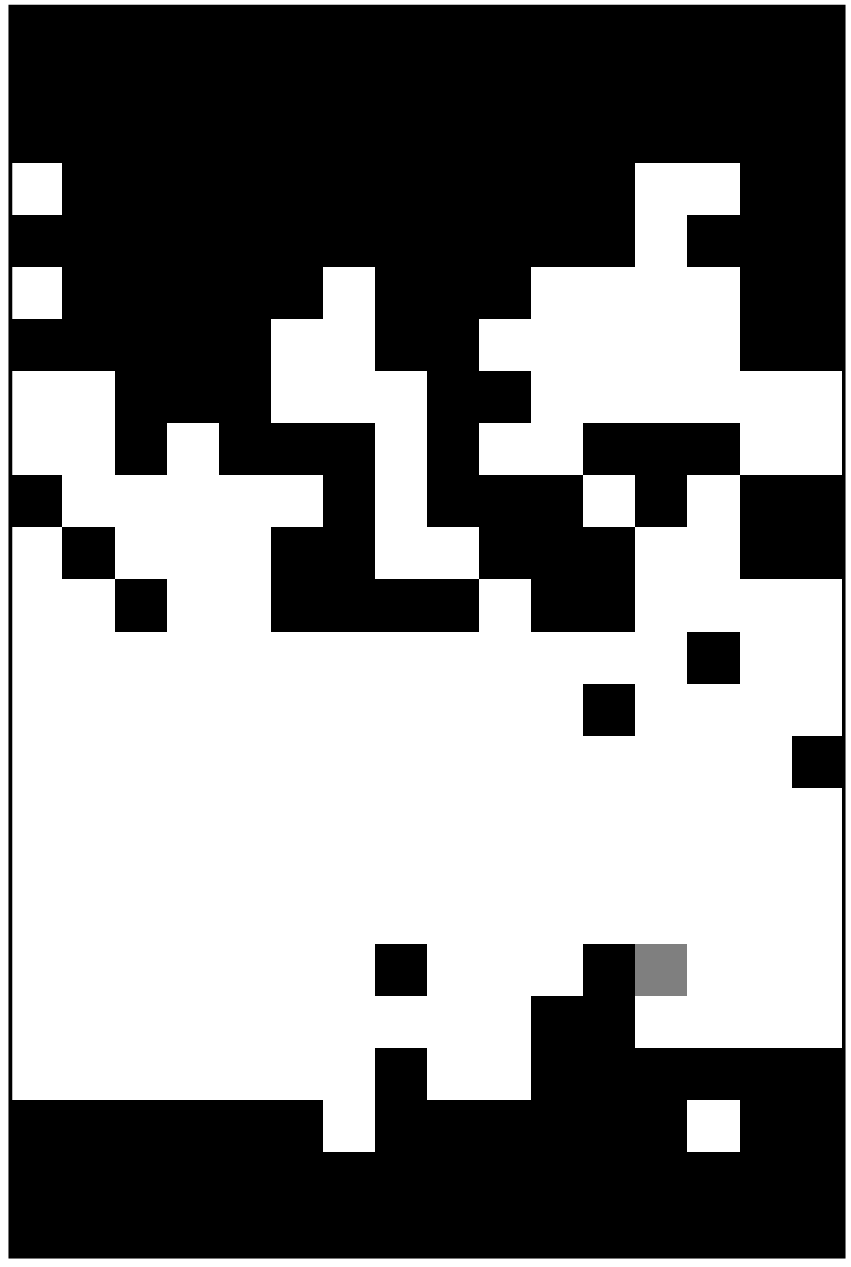}
    \begin{center} d) \end{center}
  \end{minipage}
  \hfill \hfill
\begin{minipage}{0.16\textwidth}
  \epsfxsize = \textwidth \epsfysize = 1.5\textwidth \hfill
  \epsfbox{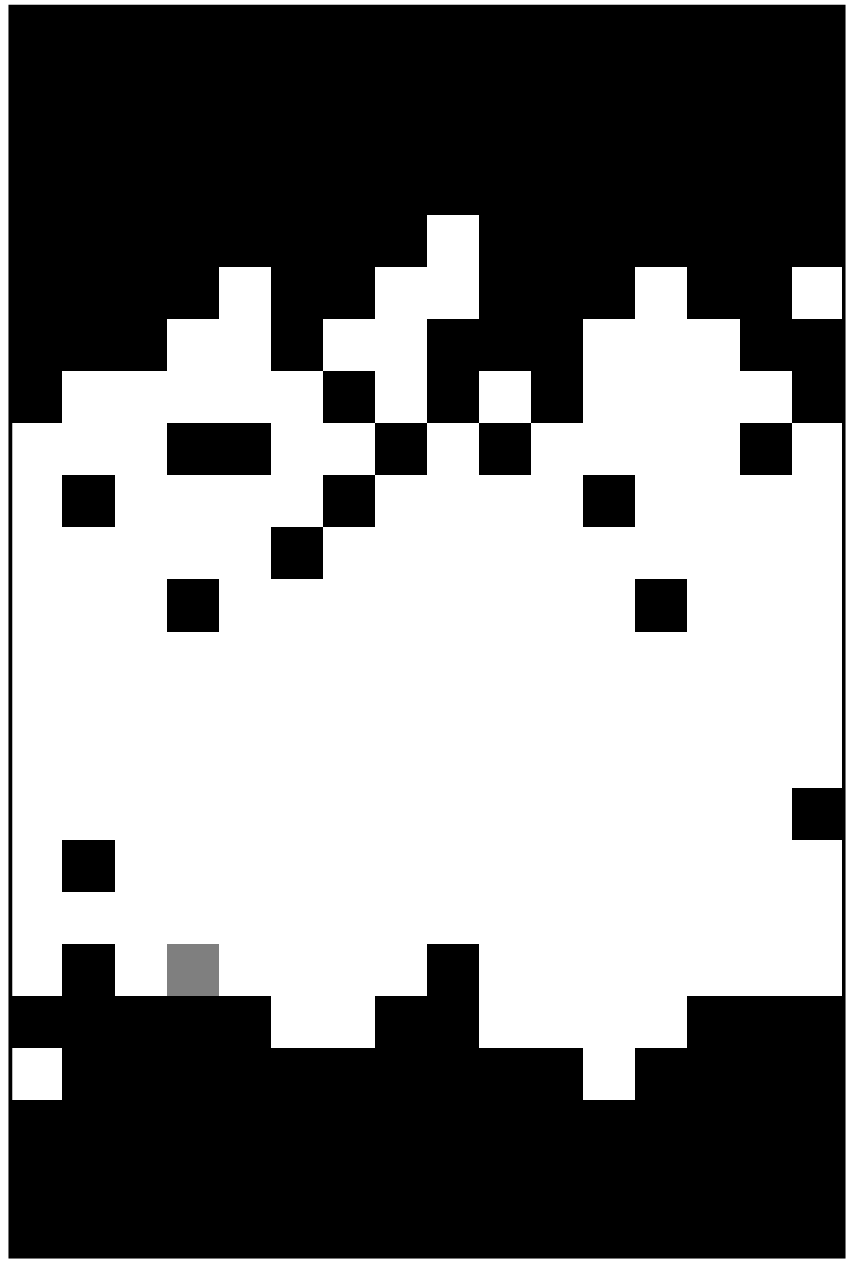}
    \begin{center} e) \end{center}
  \end{minipage}
  \hfill \hfill
\begin{minipage}{0.16\textwidth}
  \epsfxsize = \textwidth \epsfysize = 1.5\textwidth \hfill
  \epsfbox{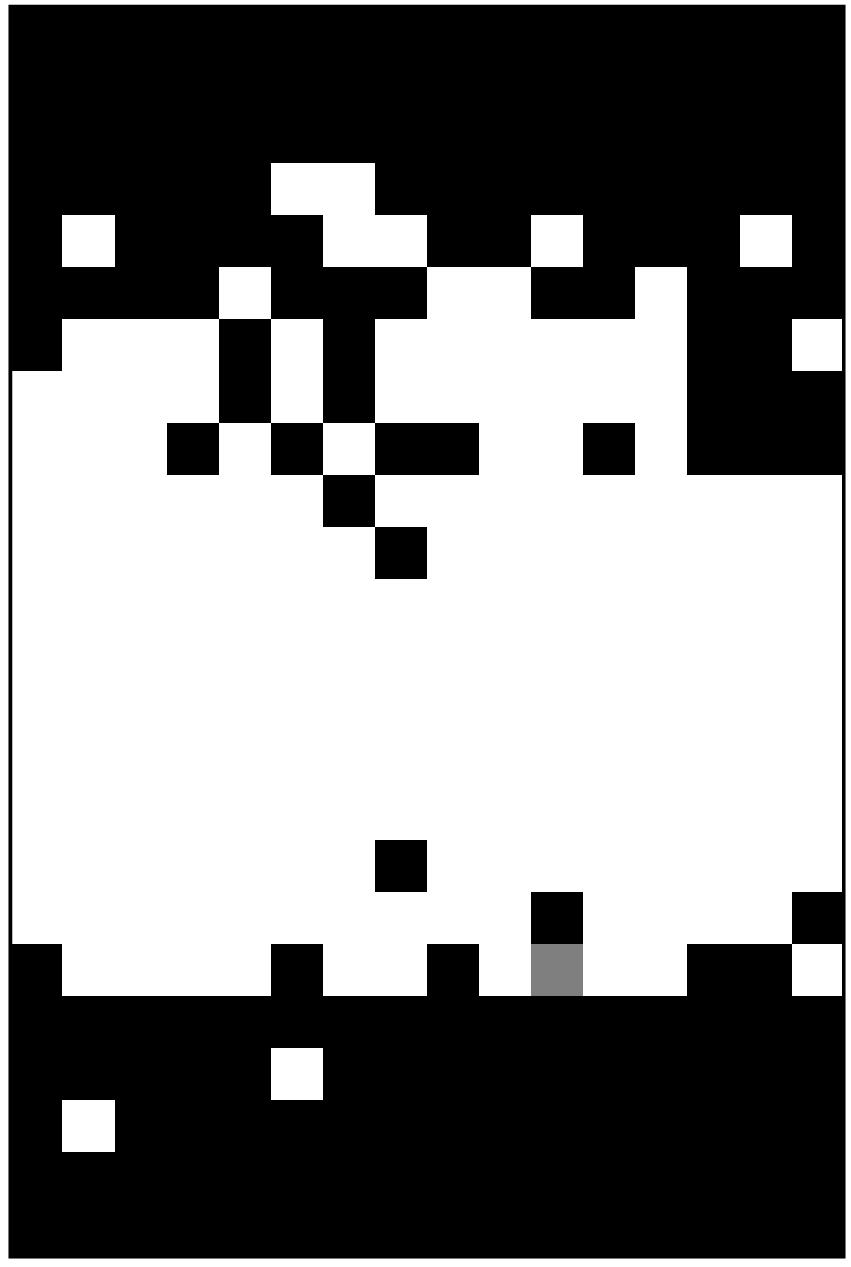}
    \begin{center} f) \end{center}
  \end{minipage}
  \hfill \hfill 
\vspace{0.03\textwidth}
\caption{Snapshots of an $L_{x}\times L_{y}$ system with $L_{x}=16$ and $%
L_{y}=24$ at $E=0.8$, at different numbers of MCS: a)$1$, b)$10^{3}$, c)$%
10^{4}$, d)$10^{5}$, e)$10^{6}$, f)$10^{7}$. The initially disordered system
undergoes a charge segregation. In the ordered steady-state configuration
the two oppositely charged particles are separated by two different
interfaces. Minus particles are colored black, plus white and the hole is
gray. $E$--field and $y$--direction point upwards.}
\label{6latticepic}
\end{figure}
\noindent
due to the periodic boundary conditions, a second
interface must also form. After $10^{5}$ MCS (Fig.~1d) the segregation of
charges, and hence the two interfaces, are quite apparent. Due to the drive,
the hole moves rapidly to the top (bottom) in regions of predominantly
negative (positive) charge. Thus, it tends to remain near the interface
which separates positive particles on the top from negative ones on the
bottom (the lower interface in Figs.~1d-f). In contrast, it is rapidly
driven away from the opposite interface. We will refer to the former
(latter) interface as the ``downstream'' (``upstream'') one. The two
interfaces are well separated, for this choice of parameters, and exhibit
rather different morphologies: The downstream interface is quite sharp,
while the upstream interface appears to be much more diffuse. To increase
the degree of order in the system, the hole has to travel to the upstream
interface before it can move another charge to a preferred position. Since
this requires a series of {\em field-suppressed} jumps, the approach to the
final steady state is very slow. As the ordered domain surrounding the
downstream interface grows, the hole becomes strongly localized. A
quantitative analysis of this ordering process will be provided elsewhere 
\cite{dynamics}.

A picture of a typical steady-state configuration is shown in Fig.~1f. To
characterize these structures, we investigate three characteristic
quantities: the order parameter $Q_{L}$, which provides a global measure of
order, as well as the average hole and charge density profiles which carry
more detailed information about ordered configurations. All three of them
are easily computed within the mean-field theory, as we shall presently see.
Since the spatial inhomogeneities form along the $y$--direction, the system
size $L_{x}$ is not expected to play an important role. Simulations confirm
this, provided the aspect ratio $L_{x}/L_{y}$ does not exceed a certain
threshold value which is at least $6$ in our case. For larger aspect ratios,
strip configurations with nonzero winding number may begin to form,
introducing an $L_{x}$-dependence into the problem~\cite{bassler}. These,
however, are not the subject of the present study.

\subsection{The Order Parameter}

We begin by calculating the order parameter $Q_{L}$ in the mean-field
approximation. Starting with the definition (\ref{ql}), we first express it
within the continuum theory. Clearly, the summation over the sites in the $y$%
--direction should be replaced by an integration. Using the rescaled
variable $z=y/L_{y}$ and exploiting the symmetry of the profiles around $z=%
\frac{1}{2}$, we obtain: 
\begin{equation}
Q_{L}=\frac{2}{m}\int_{0}^{1/2}\,\left[ \psi (z)\right] ^{2}\,dz=\frac{2}{%
m\,\epsilon ^{2}}\int_{0}^{1/2}\frac{\chi ^{\prime }{}^{2}}{\chi ^{2}}%
\,dz\quad .  \label{qlmft}
\end{equation}
In the last equality, we have recast $\psi $ in terms of $\chi $. To
proceed, we change integration variable, from $z$ to $\chi $. The limits of
the integral are transformed to $\chi (0)=\chi _{+}$ and $\chi (\frac{1}{2}%
)=\chi _{-}$, where we recall that these are zeroes of $U-V(\chi )$. With $%
\chi ^{\prime }=$ $\sqrt{(2j\epsilon ^{2}/3)\,(\chi _{+}-\chi )(\chi -\chi
_{-})(\chi -\chi _{1})}$, we find 
\begin{equation}
Q_{L}=\frac{4j}{3m}(Q_{1}+Q_{2}+Q_{3}+Q_{4})
\end{equation}
where we have introduced 
\begin{eqnarray}
Q_{1} &=&-\int_{\chi _{-}}^{\chi _{+}}\frac{\chi }{\chi ^{\prime }}d\chi
\label{q1} \\
Q_{2} &=&(\chi _{1}\,\chi _{-}\chi _{+})\int_{\chi _{-}}^{\chi _{+}}\frac{1}{%
\chi ^{2}\chi ^{\prime }}d\chi  \label{q2} \\
Q_{3} &=&(\chi _{1}+\chi _{-}+\chi _{+})\int_{\chi _{-}}^{\chi _{+}}\frac{1}{%
\chi ^{\prime }}d\chi  \label{q3} \\
Q_{4} &=&-(\chi _{+}\chi _{-}+\chi _{+}\chi _{1}+\chi _{-}\chi
_{1})\int_{\chi _{-}}^{\chi _{+}}\frac{1}{\chi \chi ^{\prime }}d\chi \quad .
\label{q4}
\end{eqnarray}
The last two integrals are evaluated easily, giving $1/2$ and $(1-m)/2$. The
first two integrals can be reduced to complete elliptical integrals of the
first, second and third kind.

The resulting expressions can be expressed in more compact form, using the
parameters $m$, $p$ and $R$ (see Eqns~(\ref{p}) and (\ref{r})). For that
purpose, it is helpful to write the three roots in terms of $p$ and $R$: 
\begin{eqnarray}
\chi _{1} &=&\frac{2\left( 1+R\,(p-2)\right) }{1-R^{2}\,(1-p+p^{2})} \\
\chi _{-} &=&\frac{2\left( 1+R\,(1-2p)\right) }{1-R^{2}\,(1-p+p^{2})} \\
\chi _{+} &=&\frac{2\left( 1+R\,(1+p)\right) }{1-R^{2}\,(1-p+p^{2})}
\end{eqnarray}

Then, we invoke Eqn~(\ref{1-4j}) to replace the current $j$ and Eqn~(\ref
{1-m}) to eliminate the elliptic integral of the third kind. Collecting, we
obtain $Q_{L}$: 
\begin{equation}
Q_{L}=1-\frac{1}{2m}\left\{ R(p-2)+1+3R\frac{E(p)}{K(p)}\right\} .
\label{mfeq}
\end{equation}
$E$ and $K$ are the complete elliptical integrals of the first and second
kind \cite{abramowitz}. According to Eqns~(\ref{r}) and (\ref{1-m}), $Q_{L}$
is a function of $p$ and $\epsilon $ only which can be generated
parametrically in $p$.

So far, our discussion is valid for arbitrary mass $m$. Let us now consider
the case of a single vacancy, namely $m=1-1/(L_{x}L_{y})$ which corresponds
to a system near complete filling. Considering only the leading terms in an
expansion in powers of $\delta \equiv 1/(L_{x}L_{y})$, the left hand side of
Eqn~(\ref{1-m}) is just $\delta $. As a consequence, the factor $%
1-R^{2}(1-p+p^{2})$ on the right hand side is $O(\delta )$, and so is the
current $j$, given by Eqn (\ref{1-4j}). Recalling that the upper limit $%
p_{0} $ of the $p$ range is defined by the condition: $j(\epsilon
,p_{0}(\epsilon ))=0$, we conclude that $p=p_{0}+O(\delta )$ for our case.
Tracking the effect of the finite-size corrections through our preceding
calculations, we find that $\chi _{1}$, $\chi _{+}$ and $\chi _{-}$ are all $%
O(L_{x}L_{y})$, by virtue of their common denominator. To leading order, the
hole density $\phi =1/\chi $ is therefore $O(\delta )$ as one might have
anticipated! In contrast, the charge density is $O(1)$, due to Eqn (\ref
{psid}). Since $\delta $ is very small in our study, all but the leading
terms will be neglected in the following. Then, $Q_{L}$ becomes a function
of $\epsilon $ {\em alone}! This prediction is easily checked by Monte Carlo
simulations.

In this spirit, we invoke Eqn~(\ref{1-4j}) for $j=0$ and rewrite it as $%
\epsilon ^{2}=[4K(p_{0})]^{2}(1-p_{0}+p_{0}^{2})$. Now, $\epsilon $ can be
computed numerically for a set of discrete values of $p$ in the interval $%
[0,1]$. The values of $Q_{L}(\epsilon ,m=1)$, derived in this way, are shown
as the solid theoretical curve in Fig.~2.

For large $\epsilon \gg 1$, the approximations $R\simeq \sqrt{1-4j}$ and $%
p\simeq 1$ are valid. Within the same approximation, $j$ can be replaced by $%
j=\exp (-m\epsilon /2)$ \cite{vilfan} which is vanishingly small. With $m=1$%
, $Q_{L}$ simplifies to 
\begin{equation}
Q_{L}=1-\frac{6}{\epsilon }  \label{qappr}
\end{equation}
This gives rise to the dashed curve in Fig.~2. Comparing this approximation
to the exact mean-field result, we see that both expressions are
indistinguishable for $\epsilon >18$. For smaller values of $\epsilon $, the
approximation underestimates the order parameter slightly.

Turning to simulation results, we first test the expected scaling in $%
\epsilon $. Figure 2 shows data for the order parameter $Q_{L}$, for different
square systems (ranging from $20\times 20$ to $35\times 35$) and different
electric fields ($E=0.2$ to $1.0$), plotted versus $\epsilon $. Each data
point is an average over 30 -- 50 samples. The size of the error bars is
about $0.05$ units.
\begin{figure}[tbp]
\input epsf
\begin{center}
\begin{minipage}{0.45\textwidth}
  \epsfxsize = \textwidth \epsfysize = .75\textwidth \hfill
  \epsfbox{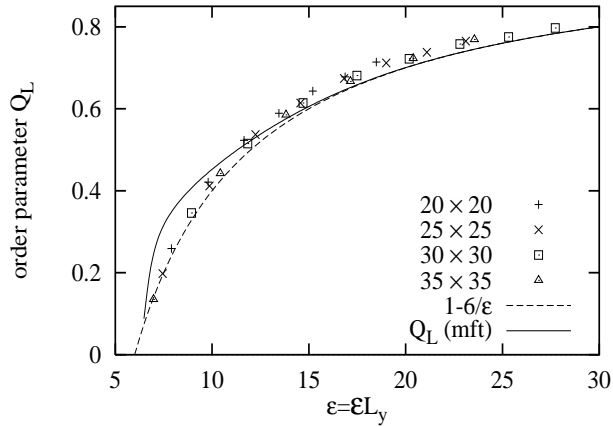} 
    \vspace{-.2cm}
\end{minipage}
\end{center}
\caption{Plot of order parameter vs $\epsilon $ for different square system
square sizes ($20\leq L_{x}=L_{y}\leq 35$) and different electric fields ($%
0.2\leq E\leq 1.0$). The upper line shows the result of $Q_{L}$ from mean
field theory, the lower line is its approximation $1-6/\epsilon $.}
\label{scalingq}
\end{figure} 
Within the accuracy of our data, all points lie on the same curve,
corresponding to the scaling function. The latter appears to tend towards
zero for $\epsilon \lesssim 6$. This is consistent with the stability limit
of the inhomogeneous solutions, Eqn~(\ref{mh}), which implies that for $%
m\approx 1$, an inhomogeneous solution can exist only if $\epsilon >2\pi $.
Once the transition to the homogeneous phase has occurred, the order
parameter is of the order of $1/L_{x}\leq 0.05$. With increasing $\epsilon $
the order parameter approaches its upper limit, i.e., $1$.

It is remarkable, however, that the large $\epsilon $ approximation produces
a better fit to the simulation data than the exact mean-field result,
especially for $\epsilon <10$. It is conceivable that the intrinsic errors
of the mean-field theory approach are partially compensated by the large $%
\epsilon $ limit. Further studies are required to test this possibility.
Focusing on the region $\epsilon >15$, the simulation results all lie about $%
0.02$ units above the theoretical curve. While these deviations are within
the error bars of the data, they are too systematic to be ignored. Closer
scrutiny reveals that the results of the large system sizes tend to be
closer to the theoretical curves than those for small system sizes, which
indicates that the differences between simulation and mean-field results are
at least partly due to finite size effects. We will return to this question
at the end of Section III.C.

\subsection{Charge and Hole Density Profiles\label%
{sectionscalingchargeandholedensity}}

While the order parameter carries only global information about spatial
inhomogeneities in the system, the charge and hole density profiles retain
far more detail, allowing us to distinguish the oppositely charged domains
and their interfaces. Based on the mean-field theory, we expect these
densities to satisfy scaling in $\epsilon $ and $z$. This is borne out by
the simulation results which are presented in this section.

In order to exhibit the scaling of the densities, four different parameter
sets ($E$, $L_{y}$) are simulated, generating $100$ samples for each. The
system length $L_{y}$ and the electric field $E$ are varied in such a way as
to keep the parameter $\epsilon $ constant at $18.24$. To avoid unwanted
cancellations, we shift the maximum of the hole density in each run to $z=0$
before averaging. The charge profiles are shifted accordingly. Thus, $z$
covers the interval $(-0.5,0.5)$, and the ``downstream'' interface is
centered at the origin. In addition, we normalize the hole profile in such a
way that all profiles enclose the same area.

A comment on this normalization is in order. Recalling the constraint on the
total density, we have $1-m=1/(L_{x}L_{y})=(1/L_{y})\int_{0}^{L_{y}}\phi
(y)dy$ for a single hole. Thus, we have $1=L_{x}\int_{0}^{L_{y}}\phi (y)dy$ $%
=L_{x}L_{y}\int_{0}^{1}\phi (z)dz$ so that $L_{x}\phi (y)$ can be
interpreted as the {\em probability density} \/for finding the hole
in row $y$. Similarly, $L_{x}L_{y}\phi (z)$ is the probability density for
finding the hole at position $z$. Thus, {\em normalized} \/plots for the
hole density show the associated probability density, and the area under
each curve is just $1$. Moreover, since $\phi (z)=O(1/(L_{x}L_{y}))$,
according to the finite-size analysis in Section IIIA, the {\em normalized}
quantity depends on $z$ and $\epsilon $ alone. No such normalization is
required for the charge density, since it is already of $O(1)$ in the system
size.

To test for the anticipated scaling in $\epsilon $ and $z$, Monte Carlo data
for the (normalized) hole density profile are presented in Fig.~3a, and the
charge density profile is shown in Fig.~3b. Since all data points collapse
onto the same characteristic scaling curve for hole and charge profiles,
respectively, the theoretical prediction is clearly confirmed.

\begin{figure}[tbp]
 \input epsf  
\begin{minipage}{0.45\textwidth}
  \epsfxsize = \textwidth \epsfysize = .75\textwidth \hfill
  \epsfbox{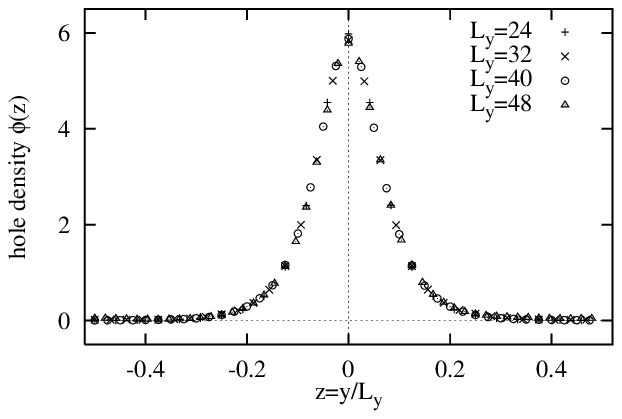} 
    \vspace{-.6cm}
    \begin{center} a) \end{center}
\end{minipage}
\hfill
\begin{minipage}{0.45\textwidth}
  \epsfxsize = \textwidth \epsfysize = .75\textwidth \hfill
  \epsfbox{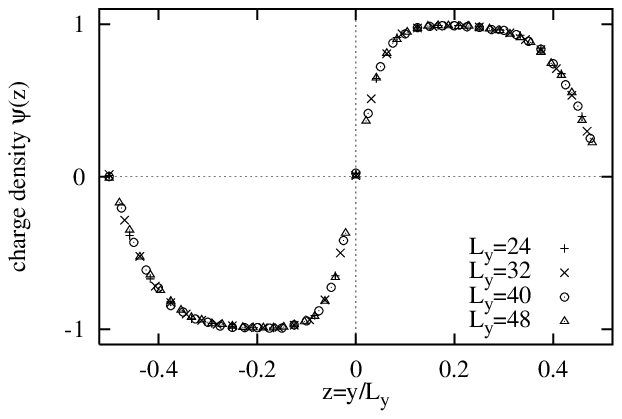} 
    \vspace{-.6cm}
    \begin{center} b) \end{center}
\end{minipage}
\vspace{.4cm}
\caption{Scaling plot of the hole (a) and charge (b) densities. $L_{y}$ and $%
E$ vary such that $\epsilon =18.24$ remains constant. $L_{x}=16$ is fixed.}
\label{ely18hd}
\end{figure}

Beyond demonstrating scaling, these plots provide a more quantitative
characterization of the spatial structures in the system. Since $\epsilon $
here is the same as in Figs.~1a-f, Fig.~3b shows the associated steady-state
charge density profiles. We can see clearly that the particles are ordered
in two regions, filling the whole system. Each of these regions consists
essentially of one species. They are separated by two interfaces. The
maximum of the hole density lies at the center ($z=0$) of the much sharper
downstream interface where the field tends to localize the hole, while the
minimum of the hole density marks the more diffuse upstream interface.\label%
{sectionsizefielddependence}

To explore the size- and field-dependence of our system further, it is
interesting to vary the system length $L_{y}$ and the electric field $E$
independently, {\em not} keeping $\epsilon $ constant. Of course, given the
excellent data collapse of Figs.~3 and 4, we cannot expect global scaling
over the whole $y$-range. We will see, however, that certain {\em regions}
of the profiles, centered on the two interfaces, still scale.

We first report simulations at constant electric field $E=0.8$ and
transverse size $L_{x}=16$, increasing the longitudinal system size $L_{y}$
from $20$ to $32$ in steps of $4$. The (normalized) hole densities observed
in these simulations are summarized in Fig.~4a, plotted vs $y$ rather than $%
z=y/L_{y}$.

\begin{figure}[tbp]
\input epsf
\begin{minipage}{0.45\textwidth}
  \epsfxsize = \textwidth \epsfysize = .75\textwidth \hfill
  \epsfbox{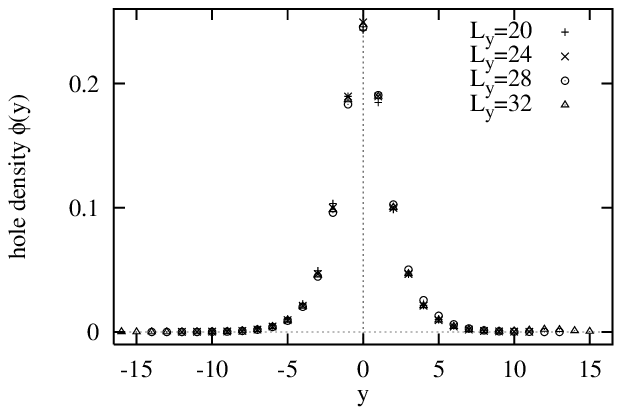} 
    \vspace{-.6cm}
    \begin{center} a) \end{center}
\end{minipage}
\hfill
\begin{minipage}{0.45\textwidth}
  \epsfxsize = \textwidth \epsfysize = .75\textwidth \hfill
  \epsfbox{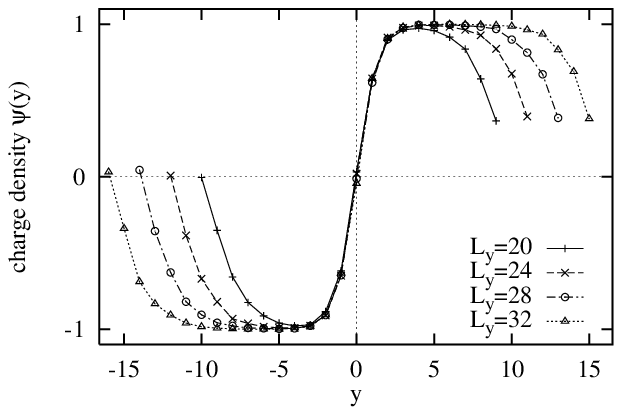} 
    \vspace{-.6cm}
    \begin{center} b) \end{center}
\end{minipage}
\vspace{.4cm}
\caption{Plot of the hole (a) and charge (b) densities for a range of $L_{y}$%
. $L_{x}=16$ and $E=0.8$ are constant.}
\label{lyde08hd}
\end{figure}

We observe that the graphs associated with different $L_{y}$ span different
ranges of $y$, but are otherwise essentially indistinguishable in the
central region. Thus, the width and the maximum of the hole density, and
hence the width of the ``downstream'' interface, are not affected by changes
in the longitudinal system size when plotted vs the $y$--variable. We
conclude therefore that the characteristics of this interface are controlled
by the electric field alone (cf.\ next section).

This behavior is also borne out by the charge density profiles, Fig.~4b.
According to Eqn (\ref{psid}), the steady-state charge and hole density
profiles are related via $\psi (y)=-\phi ^{\prime }(y)/({\cal E}\phi (y))$.
Thus, the charge densities near the ``downstream'' interface should also be
independent of $L_{y}$, in agreement with Fig.~4b. On the other hand, the
regions of nearly constant charge density must broaden, to reflect the
increasing system size. Thus, the profiles do not collapse at the edges of
the plot. However, the similarity of their form near $y=\pm L_{y}/2$
suggests that the {\em upstream} interface might scale also, provided the
profiles are shifted appropriately. This is indeed confirmed by the
simulations (cf. Fig.~5b below). Thus, the slopes and widths of the
profiles, near {\em both} interfaces, are independent of system size. The
remaining effect of $L_{y}$ is very simple and can be observed in Fig.~4b:
Outside the interfacial regions, the charge densities saturate very rapidly
at $\pm 1$, and these saturated regions expand or contract to accommodate
the selected system size.

It is now quite apparent how the profiles should scale if $L_{y}$ remains
fixed and $E$ is varied instead. Since the interfacial regions are
independent of $L_{y}$, but scale in $z$ and $\epsilon $, they must depend
on $y$ through the combination ${\cal E}\,y$. To check this conjecture, we
fix the system size at $L_{x}\times L_{y}=16\times 24$, while the electric
field increases from $0.4$ to $1.2$ in steps of $0.2$. Fig.~5a, 
shows the
(normalized) hole density profile plotted vs the scaling variable ${\cal E}y$.
 The data collapse in the interfacial region is excellent, except for the
smallest field $E=0.4$. This value of $E$, however, is rather close to the
transition line where the mean-field theory is likely to break down.
Focusing on the larger $E$'s, it is apparent that the {\em width} \/of the
downstream interface scales as $1/{\cal E}$. Turning to the charge
densities, Eqn (\ref{psid}) implies that $\psi (y)$ is also a function of $%
{\cal E}y$ near the downstream interface. This is indeed confirmed by
simulations. To illustrate the scaling of the {\em upstream} interface, we
present Fig.~5b: Here, all profiles have been shifted by $L_{y}/2$, in order
to center the upstream interface at the origin. Clearly, this interface also
scales in the variable ${\cal E}\,y$. For completeness, we note that the
only profile that does not reach saturation, is the one for the smallest $E$%
, since this value is quite close to the phase transition.

Let us summarize the key findings of the simulations. First, the data for
each profile collapse onto a single, {\em global} scaling curve if plotted
as a function of $z=y/L_{y}$ at constant $\epsilon $. Moreover, focusing
only on the interfacial (as opposed to the saturation) region, we find that 
{\em both} \/interfaces are independent of $L_{y}$ and that their widths
scale as $1/{\cal E}$, provided we are not too close to the transition to
the homogeneous phase.
\nopagebreak  
In the next section, we will consider these findings 
in the light of our mean-field theory.

\begin{figure}[tbp]
\input epsf
\begin{minipage}{0.45\textwidth}
  \epsfxsize = \textwidth \epsfysize = .75\textwidth \hfill
  \epsfbox{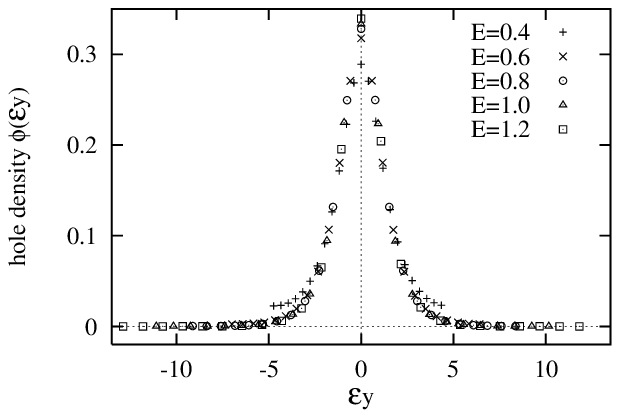} 
    \vspace{-.6cm}
    \begin{center} a) \end{center}
\end{minipage}
\hfill
\begin{minipage}{0.45\textwidth}
   \epsfxsize = \textwidth \epsfysize = .75\textwidth \hfill
  \epsfbox{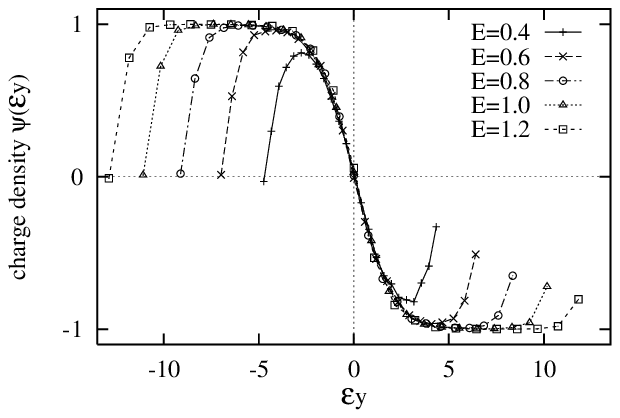} 
    \vspace{-.6cm}
    \begin{center} b) \end{center}
\end{minipage}
\vspace{.4cm}
\caption{Plot of the hole (a) and charge (b) densities for a range of $E$,
vs ${\cal E}y$. $L_{x}=16$. Note that in (b) $y=0$ corresponds to the {\em %
minimum} of the hole density (upstream interface).}
\end{figure}

\subsection{Independent Interface Approximation}

In the following, we present an analytical description of the interfaces
which is then tested by detailed Monte Carlo simulations. We will focus in
particular on the charge density since it directly determines the order
parameter. First, we invoke the ``large $\epsilon $'' approximation~\cite
{vilfan} to describe the upstream interface. A complementary approximation,
involving a different version of the large $\epsilon $ limit, is then
developed to describe the other (downstream) interface. The key assumption
here is that the two interfaces are {\em independent} from one another,
i.e., they are separated by sufficiently large ``saturation'' regions:
regions which are entirely filled by either positive or negative charges, so
that $\psi =\pm 1$ there. Such profiles result provided $\epsilon \gtrsim 18$%
, as demonstrated by Figs.~3b, 4b, and 5b. The associated hole densities are
approximately zero except near the downstream ($z=0$) interface (cf.
Figs.~3a, 4a and 5a).

Returning to the large $\epsilon $ limit of our mean-field theory, we recall
that $\epsilon \gg 1$ is equivalent to $p\rightarrow 1$. In this limit, we
may replace the Jacobian elliptic function sn by $\tanh $~\cite{vilfan}. In
practice, this is already a good approximation for $\epsilon \ge 15$. It is
easy to check that this results in (mean-field) profiles with vanishing hole
densities near the upstream interface and saturated charge densities between
the interfacial regions. Thus, this limit is consistent with our assumption
of ``independent'' interfaces.

To describe the upstream interface, we start from Eqn~(\ref{chiappr}) for
large $\epsilon $: 
\begin{equation}
\chi (z)=\chi _{+}-(\chi _{+}-\chi _{-})\tanh ^{2}(\epsilon z/2)
\label{approxchi}
\end{equation}
Note that, due to the symmetry of $\chi $, this equation holds for the
interval $(-1/2,1/2)$. At $z=0$, $\chi $ takes its maximum, so that this is
a good approximation for the upstream interface where $\phi =1/\chi $ is
minimal. The largest deviation from the exact mean-field solution occurs at
the boundaries, i.e., near the downstream interface, since this
approximation violates the periodic boundary conditions: $\chi ^{\prime
}(-1/2)\neq \chi ^{\prime }(1/2)$. The current is exponentially suppressed
for large $\epsilon $, i.e., $j\cong 6\,e^{-m\epsilon /2}$, and $\chi _{+}$
and $\chi _{-}$ can be expressed in terms of $j$, namely, $\chi _{+}-\chi
_{-}\cong \frac{3}{2j}\sqrt{1-4j}$ and $\chi _{+}\cong \frac{1}{2j}(1+2\sqrt{%
1-4j})$ \cite{vilfan}. The hole and charge densities are now easily derived.
In particular, recalling that $y=L_{y}z$, we can already read off the width $%
\xi _{u}$ of the upstream interfacial region: $\xi _{u}=2L_{y}/\epsilon =2/%
{\cal E}$ which is consistent with the data. More specifically, we can
compute the charge density from Eqns (\ref{psid}) and (\ref{approxchi}).
Neglecting terms of $O(j)$, we find 
\begin{equation}
\psi (y)=-\tanh ({\cal E}\,y/2).  \label{uinterf}
\end{equation}
At the boundaries, Eqn~(\ref{uinterf}) results in $\psi (\pm
L_{y}/2)\rightarrow \mp 1$ in the large $\epsilon $ limit, which confirms
that this approximation violates the boundary conditions. However, it does
describe the interfacial region near $y=0$ very well. Rather than quoting
the hole density explicitly, we only note that it is very small near the
origin, namely $O(j)$.

In order to capture the downstream interface, we introduce another method.
Since $j\cong 0$ to excellent accuracy for large $\epsilon $ \cite{vilfan},
we return to the mean-field equations, (\ref{MFT}), and integrate them,
setting {\em both} \/integration constants, i.e., hole {\em and} \/charge
current, to zero. This is actually an equilibrium approximation, as we shall
discuss below. The simulation data suggest the boundary conditions: $\phi
(\pm L_{y}/2)\simeq 0$ and $\psi (\pm L_{y}/2)\simeq \pm 1$. In this
approximation, the downstream interface, corresponding to the {\em maximum}
of the hole density, is localized at the origin. Written in terms of the
variable $y$, Eqn. (\ref{de1}) for $\chi $ simplifies to 
\begin{equation}
\chi ^{\prime \prime }(y)/{\cal E}^{2}=\chi (y)-1.  \label{de2}
\end{equation}
This is easily solved, subject to the specified boundary conditions: 
\begin{equation}
\chi (y)=1+c\cosh \,({\cal E}y)\quad .
\end{equation}
To ensure that the hole density is strictly positive, we demand $c>0$. This
constant can be determined explicitly from the mass constraint, namely $%
1=L_{x}\int_{-L_{y}/2}^{L_{y}/2}\phi (y)\,dy$, whence 
\begin{equation}
2\frac{L_{x}L_{y}}{\epsilon \sqrt{c^{2}-1}}\arccos \left( \frac{1}{c}\right)
=1  \label{detceqn}
\end{equation}
in the large $\epsilon $ limit. In our simulations, $L_{x}$ is at least $16$
and ${\cal E}$ at most $2$, so that $c>24$ follows. Thus, we can expand Eqn (%
\ref{detceqn}) for large $c$, resulting in 
\begin{equation}
c\simeq \frac{L_{x}L_{y}}{\epsilon }\pi  \label{eqc}
\end{equation}
In fact, this approximation is already very good for $c>4$. Next, we compute
the charge density, 
\begin{equation}
\psi (y)=\frac{\sinh ({\cal E}y)}{\cosh ({\cal E}y)+c^{-1}}\simeq \tanh (%
{\cal E}\,y)\quad .  \label{dinterf}
\end{equation}
The last approximation is very accurate since $c>24$. Again, we can read off
the width of the interfacial region, $\xi _{d}=1/{\cal E}$. Similar to the
downstream interface, the width scales with $1/{\cal E}$, in agreement with
the data. Intriguingly, however, our approximation is capable of reproducing
the observation that the downstream interface is {\em narrower} \/than the
upstream one. Whether the measured widths differ by a simple factor of $2$,
as predicted by our calculation, awaits a more quantitative comparison with
Monte Carlo data.

In contrast to the upstream interface, $\phi $ is nontrivial here: 
\begin{equation}
\phi (y)=\frac{1}{1+c\cosh ({\cal E}y)}\simeq \frac{1}{c\,\cosh ({\cal E}y)}
\label{dinterfphi}
\end{equation}
confirming the width of the downstream interface $\xi _{d}=1/{\cal E}$. Away
from the origin, the hole density again decays very rapidly, to match with
its value near the upstream interface.

Before turning our focus on a comparison of these results with computer
simulations, a last remark on the approximation of the downstream interface
is in order. Imposing brick-wall (i.e., closed) rather than periodic
boundary conditions, the approximation taken here (setting the current to
zero) becomes exact. Moreover, the brick-wall system is an {\em equilibrium}
\/one. The hole will accumulate positive (negative) charged particles at the
top (bottom) of the system, thus establishing our boundary condition $\psi
(\pm L_{y}/2)\rightarrow \pm 1$. Clearly, only one nontrivial interface
remains in this case, namely, the downstream one. In the steady state, the
bias traps the hole near this interface. This fixes the boundary condition $%
\phi (\pm L_{y}/2)\rightarrow 0$. In this sense, our approximation for the
downstream interface is equilibrium-like.

Returning to our model, we have obtained two compact equations, (\ref
{uinterf}) and (\ref{dinterf}), for the charge density. Since $\psi
(y)\simeq \pm 1$ between the interfaces, to excellent accuracy, the whole
system can be described in terms of the two interfaces, provided we match
them appropriately. As an example, Fig.~6b shows a $16\times 24$ system with 
$E=0.8$. The data points result from a Monte Carlo simulation while the
solid and dashed lines reflect our two interface approximations, Eqns~(\ref
{uinterf}) and (\ref{dinterf}), respectively. For the narrower interface
(downstream; in the center), the match is nearly perfect, while for the
wider interface (upstream; at the edges of the figure) the slope of the $%
\tanh $--function is slightly too small compared to the computer simulation.
The agreement is nevertheless remarkable.

\begin{figure}[tbp]
\input epsf
\begin{minipage}{0.45\textwidth}
  \epsfxsize = \textwidth \epsfysize = .75\textwidth \hfill
  \epsfbox{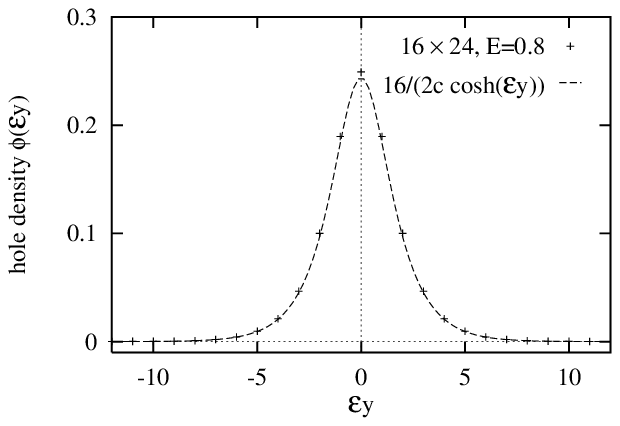} 
    \vspace{-.6cm}
    \begin{center} a) \end{center}
\end{minipage}
\hfill
\begin{minipage}{0.45\textwidth}
    \epsfxsize = \textwidth \epsfysize = .75\textwidth \hfill
  \epsfbox{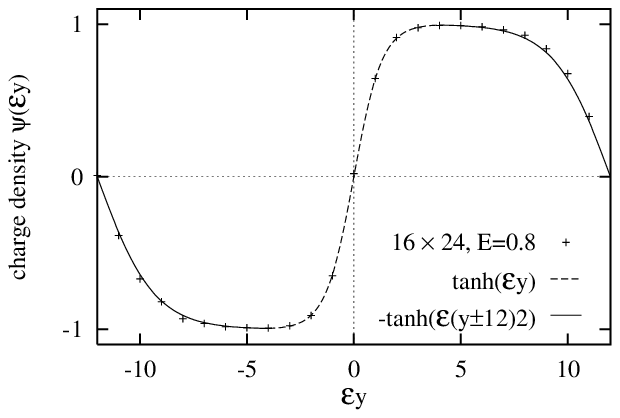} 
    \vspace{-.6cm}
    \begin{center} b) \end{center}
\end{minipage}
\vspace{.25cm}
\caption{(a) Plot of the hole density (+) for $L_{x}=16$, $L_{y}=24$ and $%
E=0.8 $. The dashed line denotes the approximation Eqn (\ref{dinterf}), for
the downstream interface, for $c=65.58$. Note that the hole density is
vanishingly small near the upstream interface.
(b) Plot of the charge density (+) for $L_{x}=16$, $L_{y}=24$ and $E=0.8 $.
The dashed and solid lines are the two interface approximations, Eqns~(\ref
{uinterf}) and (\ref{dinterf}), matched at $y=\pm 4$.}
\end{figure}

Next, we compare the approximation for the hole density with MC results.
Here, we use Eqn~(\ref{dinterfphi}) for the whole system since the hole
density is vanishingly small except in the central region of the downstream
interface. Simulation data and the analytic approximation for $\phi $ are
presented in Fig.~6a. A small quantitative discrepancy is observed at $y=0$,
i.e., the center of the downstream interface, whereas all other data points
are remarkably well reproduced by Eqn~(\ref{dinterfphi}).

Given the results for the interfaces, we finally return to the order
parameter $Q_{L}$. Here, we will see that the independent interface
approximation provides us with a very intuitive picture for the approximate
form (\ref{qappr}). Since the steady state exhibits complete order in one
region of positive and another one of negative particles, the deviation of $%
Q_{L}$ from unity originates near the interfaces. We can easily compute the
contribution to $Q_{L}$ for each interface separately, using Eqn~(\ref{qlmft}%
). The wider (upstream) interface reduces $Q_{L}$ by $4/({\cal E}L_{y})$,
while the narrower downstream interface lowers it by $2/({\cal E}L_{y})$,
resulting in a net $Q_{L}=1-6/({\cal E}L_{y})$, in agreement with Eqn~(\ref
{qappr}). Thus, this form simply tallies up the contributions of two
well-separated interfaces, while the fully saturated regions give rise to
the $1$.

While these data for the scaling of order parameter and profiles are very
convincing, the question of their range of validity must be raised. First,
we should anticipate a breakdown of mean-field theory near the onset of the
transition to the uniform state. This limits our analysis to $\epsilon
\gtrsim 6$, corresponding to, e.g., $E\gtrsim 0.25$ for a system with $%
L_{y}=24$. For larger values of $\epsilon $ (but below an upper limit to be
discussed shortly), scaling in $\epsilon $ and $z$ is observed to hold. In
order to have well-separated interfaces, we also require $\epsilon \gtrsim
15 $. Beyond this threshold, the interfacial regions of the profiles scale
very cleanly with $1/{\cal E}$.

In addition to a lower limit, there is also an upper limit for our analysis.
Recalling Eqn (\ref{curly}), the effective drive is bounded: ${\cal E}\leq 2$%
, due to the $\tanh $-function, even for very large values of the
microscopic $E$. Thus, within our mean-field theory, the interfacial widths
cannot become arbitrarily small. For example, for the narrower downstream
interface $\xi _{d}\leq 2$ in units of the lattice spacing, and $\psi (\pm
2)=\pm 0.96$, from Eqn (\ref{dinterf}). Thus, mean-field profiles require 
{\em at least} 4 lattice spacings, to interpolate between the fully
saturated regions. In contrast, {\em measured} charge density profiles for
large $E$ (e.g., $E=2.0$), jump from $-1$ to $+1$ over just two lattice
spacings! Such profiles are so sharp that our continuum limit fails to
reproduce them: they can hardly be considered smoothly varying functions. As
a result, the mean-field theory {\em underestimates} the order parameter for
large values of $E$ which explains the systematic deviations of the smaller
system sizes in Fig.~2. For example, $\epsilon =20$ in a $20\times 20$
system corresponds to $E=1.2$ where this phenomenon is already noticeable.
At a purely phenomenological level, we can extend the validity of our
mean-field description if we retain the {\em form} of our equations, (\ref
{MFT}), but replace the effective drive ${\cal E}$ by the microscopic field $%
E$ everywhere. Mathematically, this requires keeping {\em explicit} track of
the lattice constant $a$, followed by taking the {\em hydrodynamic} limit \cite {spohn},
i.e., $a\rightarrow 0$ at fixed drive, system size and mass. Since the lattice
constant appears in the rates, Eqn (\ref{transitionrate}), the effective
drive takes the form ${\cal E}=2\tanh (Ea/2)$. In the original discrete
version of Eqn (\ref{MFT}), the lattice constant appears in terms such as $%
{\cal E}\left\langle \frac{1}{2}(\phi _{x+a,y}-\phi _{x-a,y})\right\rangle $%
. In the limit of vanishing $a$, this expression simplifies to $%
Ea^{2}\partial \phi /\partial x$. Since the diffusive terms (e.g., $\nabla
^{2}\phi $) also generate a factor $a^{2}$, the latter can be absorbed into
the time scale so that we recover Eqn (\ref{MFT}), with ${\cal E}$ replaced
by $E$. Thus, all of our analytic results carry over, provided $E$ takes the
place of ${\cal E}$ everywhere \cite{comment}. With this modification, the
agreement of MC data and analytic description extends to the largest fields
studied, namely, $E=2.0$.

To some extent, even the measured profiles do not reproduce the actual
sharpness of the data fully. Since the downstream interface can form at
arbitrary location within the lattice, one should allow for {\em noninteger}
shifts, i.e., shifts between 0 and 1 ($a=1$) modulo multiples of the lattice
spacing, in order to produce accurate averaged data. This subtlety is not
accounted for in our simulations, as seen from the discussion at the end of
Section~\ref{mcsimul}. Thus, the actual interface is slightly smeared out
when we average profiles by superposing the maxima of the hole density.
Details can be found in Ref. \cite{master}. To summarize briefly, our
mean-field theory, in the form of Eqns (\ref{MFT}), gives excellent results
provided $\epsilon \gtrsim 6$ and $E\leq 1.0$. If a systematic hydrodynamic
limit is considered, the validity extends further, at least to $E\leq 2.0$.

\section{Conclusions}

In this work, we focused on the scaling behavior of ordered steady states in
a simple lattice model. A fully periodic lattice is filled with equal
numbers of positive and negative ``charges'', except a single site which
remains empty. An external ``electric'' field, applied along one of the
lattice axes, biases the motion of the particles. The dynamics is
vacancy-mediated in that only vacancy-charge exchanges are allowed. The
particles interact only through an excluded volume constraint.

This system develops spatial structures if $EL_{y}$, i.e., the product of
drive and system size along the field direction, exceeds a critical value.
Then, a charge segregated strip, oriented transverse to the field, forms
around the hole and grows until it fills the whole system. The two
oppositely charged regions are separated by two interfaces with distinct
characteristics: One interface, the ``downstream'' one, attracts the hole,
the other (upstream) repels it strongly. This asymmetry finds its origin in
the charge separation induced by the external field: while the hole moves
rapidly {\em along} the field in the negative region, its preferred
direction is reversed in the positive region.

Continuing earlier studies \cite{schmittmann1992,vilfan}, we investigate the
scaling properties of an appropriate order parameter and the hole and charge
densities, as the external control parameters $E$ and $L_{y}$ vary. The
transverse system size $L_{x}$ plays no role except in finite-size
corrections. Monte Carlo data are compared to the predictions of a
mean-field theory in which the drive appears through the effective parameter 
${\cal E}\equiv 2\tanh (E/2)$. The agreement is excellent, provided $%
\epsilon \equiv {\cal E}L_{y}\gtrsim 6$ so that we are in the ordered phase,
and $E\leq 1.0$ to maintain fairly smooth profiles. The transverse system
size $L_{x}$ plays no role except in finite-size corrections. In particular,
we can describe the charge density profiles, with remarkable accuracy, in
terms of two non-interacting interfaces, separated by perfectly ordered
regions. The interfaces themselves are determined by the drive alone,
independent of system size, and their widths scale with $1/{\cal E}$. For
fields $E>1.0$, the data show very steep slopes in the interfacial regions
which cannot be captured correctly by a naive continuum limit. Remarkably,
the mismatch between data and mean-field theory is significantly reduced if
we substitute the {\em microscopic} field $E$ for the {\em effective} ${\cal E}$ 
in the (mean-field) interface approximations. The emergence of the latter
can be understood in the limit of vanishing lattice constant. 
We should caution,
however, that this limit must also eventuelly break down since it does not 
commute with the limit $E\rightarrow \infty $. 
Since the
details of the continuum limit appear to play a key role here, it would be
interesting to analyze the {\em discrete} precursor of Eqn (\ref{MFT}). In
this case, the natural parameter should be ${\cal E}$ alone. 

Another interesting question concerns the character of the interfaces when a
finite density of vacancies is present. In this case, the downstream
interface ``splits'' into two halves, separated by an empty region. Clearly,
in addition to $E$ and $L_{y}$, the overall mass $m$ now enters the
criterion for having independent interfaces. Provided the appropriate
condition is met, however, we expect that the interfacial profiles still
depend only on ${\cal E}$: {\em local} structures appear to be controlled
entirely by the drive.

Finally, our study prepares the ground for the exploration of dynamic
phenomena in driven two-species models. Having established the scaling
properties of the {\em final} steady states, work is in progress to
investigate how they {\em develop} from random initial conditions \cite
{dynamics}. \smallskip \medskip 

\smallskip \medskip \noindent {\bf ACKNOWLEDGMENTS:}

We wish to thank R.K.P.~Zia, G.~Korniss, and Z.~Toroczkai 
for valuable discussions. Support from the US National
Science Foundation through the Division of Materials Research is gratefully
acknowledged.

\end{document}